\documentclass[aip,jap,reprint,showpacs,showkeys]{revtex4-1}
\pdfoutput=1

\usepackage[pdftex]{graphicx}
\usepackage{amsmath}
\usepackage{dcolumn}
\usepackage{bm}
\usepackage{epstopdf}
\usepackage{threeparttable}
\usepackage{adjustbox}

\begin{document}

\title{Isochronal annealing effects on local structure, crystalline fraction, and undamaged region size of radiation damage in Ga-stabilized $\delta$-Pu }

\author{D.~T.~Olive}
\affiliation{Chemical Sciences Division, Lawrence Berkeley National Laboratory, Berkeley, California 94720, USA}
\affiliation{Materials Science and Technology Division, Los Alamos National Laboratory, Los Alamos, New Mexico 87545, USA}

\author{D.~L.~Wang}
\affiliation{Nuclear Science Division, Lawrence Berkeley National Laboratory, Berkeley, California 94720, USA}
\affiliation{Department of Chemistry, University of California Berkeley, Berkeley, California 94720, USA}

\author{C.~H.~Booth}
\email[Corresponding author: ]{chbooth@lbl.gov}
\affiliation{Chemical Sciences Division, Lawrence Berkeley National Laboratory, Berkeley, California 94720, USA}

\author{E.~D.~Bauer}
\affiliation{Materials Physics and Applications Division, Los Alamos National Laboratory, Los Alamos, New Mexico 87545, USA}

\author{A.~L.~Pugmire}
\affiliation{Materials Science and Technology Division, Los Alamos National Laboratory, Los Alamos, New Mexico 87545, USA}

\author{F.~J.~Freibert}
\affiliation{Materials Science and Technology Division, Los Alamos National Laboratory, Los Alamos, New Mexico 87545, USA}

\author{S.~K.~McCall}
\affiliation{Materials Science Division, Lawrence Livermore National Laboratory, Livermore, California 94550, USA}

\author{M.~A.~Wall}
\affiliation{Materials Science Division, Lawrence Livermore National Laboratory, Livermore, California 94550, USA}

\author{P.~G.~Allen}
\affiliation{Materials Science Division, Lawrence Livermore National Laboratory, Livermore, California 94550, USA}

\hyphenation{EXAFS}

\date{\today}

\begin{abstract}
The effects on the local structure due to self-irradiation damage of Ga stabilized $\delta$-Pu stored at cryogenic temperatures have been examined using extended x-ray absorption fine structure (EXAFS) experiments. Extensive damage, seen as a loss of local order, was evident after 72 days of storage below 15 K. The effect was observed from both the Pu and Ga sites, although less pronounced around Ga. Isochronal annealing was performed on this sample to study the annealing processes that occur between cryogenic and room temperature storage conditions, where damage is mostly reversed. Damage fractions at various points along the annealing curve have been determined using an amplitude-ratio method, standard EXAFS fitting, and a spherical crystallite model, and provide information complementary to previous electrical resistivity- and susceptibility-based isochronal annealing studies. The use of a spherical crystallite model accounts for the changes in EXAFS spectra using just two parameters, namely, the crystalline fraction and the particle radius. Together, these results are discussed in terms of changes to the local structure around Ga and Pu throughout the annealing process and highlight the unusual role of Ga in the behavior of the lowest temperature anneals.     
\end{abstract}


\maketitle

\section{Introduction}
\label{intro}

Although understanding the damaging effects of self-irradiation is, in principle, solvable using molecular dynamics techniques, there are particular challenges yet to be overcome when studying such effects in materials containing plutonium, especially elemental plutonium. One challenge is obtaining appropriate potentials for Pu, a problem that is exacerbated by our restricted knowledge of crystal structures of Pu compounds and Pu's complicated involvement of 5$f$ orbitals in bonding. Related to this problem is the need to perform simulations at elevated temperature in order to limit the number of atoms required to calculate a single damage cascade event by rapid annealing of defects. While such calculations can be performed and much can be learned from them, the need for elevated temperatures and the consequent small amount of damage predicted by such calculations is difficult to observe in real experiments. In this paper, we describe local structure experiments on Ga-stabilized $\delta$-Pu that decouple the damage production and annealing processes, including using a new model that accounts for undamaged-region size effects that sheds new light on both processes and highlights the effects of the Ga atoms.

The metallurgy of plutonium is inherently complex.\cite{morss2006chemistry} Coupled with changes in chemistry and structure as a result of radioactive decay, plutonium has aptly been described as ``never at equilibrium.''\cite{hecker2008plutonium} The primary practical interest in plutonium lies with its importance for energy generation applications in both nuclear power and nuclear weapons. However, there are also connections to fundamental science, as understanding damage at the atomic level is complicated by plutonium's unique electronic and structural properties.\cite{booth2012multiconfigurational,hecker2004magic} An understanding of the electronic properties is, in fact, required for calculating the structure and thermodynamics of plutonium and its compounds from first principles.\cite{jomard2008structural,bo2008first,hernandez2014ab} Even after being studied for 75 years,\cite{seaborg1946radioactive} we are just now starting to understand the origin of some of its unique behavior.\cite{Janoschek:2015fk}

Predicting how radiation damage will alter the properties of a material, and how those property changes evolve over time requires a detailed understanding of the damage processes associated with radiation damage over many length scales. The effects of radiation damage were actually discovered before radioactivity in metamict materials,\cite{ewing2007materials} that is, minerals that had once been obviously crystalline, but through radioactive decay had become amorphous.\cite{ewing1994metamict} In the case of $\alpha$-decay, the $\alpha$ particle and the resultant recoiling nucleus deposit their energies as they travel through the crystal structure, displacing atoms from their normal positions and thereby creating vacancies and interstitials. The alterations to the crystal structure are capable of driving changes in physical dimensions as a result of void swelling, the creation of new chemical phases, and other changes to the microstructure of the material.\cite{was2007fundamentals} These physical processes drive changes in other material properties as well, including thermal conductivity,\cite{vineyard1976thermal} strength,\cite{matzke1982radiation} and electrical resistivity.\cite{wigley1965effect}

The $\alpha$-decay of Pu-239 results in two recoiling nuclei, usually an 88 keV U-235 and a 5.1 MeV $\alpha$~ particle.\cite{artna1971nuclear} According to calculations, the U travels around 12 nm, depositing about 52.7 keV into elastic collisions \cite{de1998self,wolfer2000radiation} across a 7.5 nm cascade and generating $\sim$2290 Frenkel pairs. The helium nucleus travels farther, about 11 $\mu$m, but deposits only about 12 keV into elastic collisions generated in a 1 $\mu$m cascade.\cite{de1998self,wolfer2000radiation}

Such simplistic calculations do not account for the role of finite temperature well, and in fact, annealing of interstitial/vacancy pairs and other damage is a critical process in actual materials. Looking at the resistivity of radiation damaged and quenched metals, Schilling and co-workers\cite{schilling1973recovery,schroeder1976annealing} label the various stages of recovery along the annealing curve, and describe the physical processes responsible for them: In Stage I recovery, interstitials next to their own vacancies (Frenkel pairs) are thought to recombine as soon as there is enough thermal energy to do so. As more energy becomes available, the interstitials can migrate more freely to find vacancies farther away, or find other impurity traps.\cite{schilling1973recovery} In Stage II, interstitials rearrange, and detrapping of interstitials from impurity sites occurs.\cite{schilling1973recovery} Stage III sees vacancies becoming mobile, and annihilating with any remaining interstitials,\cite{schilling1973recovery} or starting to form vacancy clusters.\cite{schroeder1976annealing} These vacancy clusters may grow larger in Stage IV, before thermally dissociating in Stage V.\cite{schroeder1976annealing}

Computer models have been used to simulate damage in materials using methods based on electronic structure calculations,\cite{korhonen1995vacancy,dudarev2013density} kinetic Monte Carlo,\cite{de1998self} and molecular dynamics,\cite{van2007classical,robinson1974computer} including $\delta$-Pu.\cite{kubota2007collision,wolfer2000radiation} However, several problems arise, including the instability of the models at low temperature, and difficulty verifying structures produced at the smallest scales. Those simulations highlight the importance of the delicate interplay between damage creation processes and recovery processes in the evolution of microstructure.

In order to separate those two processes and gain a better understanding of them, cryogenic aging can be used to allow damaged material to accumulate quickly such that the damage has local structure manifestations, followed by isochronal annealing to study the damage repair mechanisms.\cite{wigley1965effect,jacquemin1973thesis,mccall2006emergent,fluss2004temperature} Here we show the results of Pu $\mathrm{L_{III}}$-edge and Ga K-edge extended x-ray absorption fine structure (EXAFS)\cite{koningsberger1987x,bunker2010introduction} experiments taken after various stages of isochronal annealing after a cryogenic aging and storage phase of 72 days. Differences in thermal behavior at various annealing temperatures are vastly reduced by collecting nearly all the data at low temperature 15~K. In this way, differences in bond length disorder as measured by the Debye-Waller factors, $\sigma^2$, can be attributed to static bond length disorder, as opposed to differences in thermal behavior. The data for both Pu and Ga edges are fit using standard EXAFS techniques to understand changes in local coordination throughout the damage recovery process. Additionally, the data are modeled using a spherical crystallite model to parameterize the size and fraction of undamaged regions in the material, providing new insight into both the damage production and annealing mechanisms.  

\section{Experimental Details}
\label{exp}

The nominal sample composition is Pu with 4.3 $\pm$ 0.2 at.~\% Ga, with a typical grain size of 25 $\mu$m. The isotopic content of the Pu consists (by atomic \%) of about 93.4\% Pu-239, 5.95\% Pu-240, 0.172\% Pu-241, 0.138\% Pu-242, and 0.040\% Pu-238, with a 0.287\% Am-241 impurity. The isotopic mixture of plutonium in this sample undergoes $3.43 \times 10^{-5}$ $\alpha$-decays per Pu atom per year. Correcting for Ga concentration, after 72 days, the sample has been exposed to $6.48 \times 10^{-6}$ decays per atom, a measure of damage which, unlike displacements per atom (dpa), makes no assumption as to the size or annealing of the damage cascade. For comparison to other studies however, the approximate rate of damage of 1~dpa for every 10~years \cite{wolfer2000radiation} would estimate this sample was damaged to 0.02~dpa.

The sample was $\approx$ 80 $\mu$m thick, and preparation steps included vacuum annealing, electropolishing, and coating in liquid Kapton with heat cure as detailed in Ref. \citenum{booth2013self}. It was loaded into an aluminum sample holder under dry N$_2$. Kapton windows sealed with indium wire and epoxy provided optical access to the sample, which nested into two additional layers. The sample was placed in a temperature-controlled closed-cycle liquid He cryostat (Montana Instruments). The sample was annealed for 30 minutes at 375 K using the heaters in the cryostat, then chilled to below 5 K. A brief power outage resulted in the temperature rising to about 14 K partway into the experiment. Out of 72 days in cold storage, the last 46 were uninterrupted below 5 K. A final transient occurred during the move from the cryostat storage location to the beamline, with the sample warming to $\approx$15 K during the time the cryostat was unplugged. The EXAFS measurements were conducted at beamline 11-2 at the Stanford Synchrotron Radiation Lightsource (SSRL). Wavelength was selected using a Si(220) ($\phi=0$) double crystal monochromator, detuned 50\% to remove harmonics. The samples were measured in fluorescence geometry using a 100-element Ge detector (Canberra), and deadtime corrections were applied.

The EXAFS data were reduced according to standard procedures.\cite{rehr2000theoretical,koningsberger1987x} Data reduction, including summation, calibration, and error corrections, was performed using SixPack\cite{webb2005sixpack} and Athena.\cite{ravel2005athena} Error bars for EXAFS parameters were determined by taking the standard deviation of the best fit values for repeated scans, with most scans repeated four times, and none repeated fewer than three times. 

The collected spectra suffered from self-absorption effects, altering the apparent amplitudes of the EXAFS oscillations. Self-absorption corrections were applied to the data,\cite{booth2005improved} bringing the coordination numbers in the fully annealed sample to the values expected from undamaged bulk samples.

Isochronal annealing was carried out by warming the sample to a series of higher temperatures ($T_A=35$, 45, 55, 65, 75, 95, 105, 115, 125, 135, 156, 200, and 300~K), holding at that temperature for 5 minutes, then cooling back down to 15 K for EXAFS measurements. Heating to and cooling from the $T_A=35$~K took less than 10 minutes, rising to around 2 hours each direction for the $T_A=300$~K. Although the EXAFS for each data point for each edge took approximately 1 hour to collect, the measurements lasted over 2.5 days of beamtime due to extra time spent heating and cooling. We note unlike electrical resistivity measurements\cite{fluss2004temperature} in which the isochronal annealing curve is run twice, first to measure damage accumulated and then again to measure residual resistivity, the EXAFS isochronal annealing curve is only run one time, and in that sense is much closer to the protocol used in magnetic susceptibility isochronal annealing experiments.\cite{mccall2006emergent} 

The data presented here are improved relative to a previous EXAFS annealing study on the same 4.3 at.~\% Ga $\delta$-Pu sample\cite{booth2013self} in several respects. Most importantly, the present data were all collected at $T = 15$~K, whereas in Ref.~\onlinecite{booth2013self}, the data were collected at $T=T_A$ and therefore the thermal broadening of the EXAFS signal had to be subtracted to isolate the annealing effect. In addition, no data were collected between about 50~K and 140~K in the previous study. 

\section{Fitting Methodologies}
\label{meth}

EXAFS is a function of photoelectron wave number,
\begin{equation}
\label{eq:k}k=\sqrt{\frac{2m_e}{\hbar^2}(E-E_0)},
\end{equation}
where $m_e$ is the electron mass, and $E_0$ is the photoelectron threshold energy, and it is governed by the so-called EXAFS equation:
\begin{equation}
\label{eq:exafs}\chi(k) = \sum_{i} \frac{N_{i} S_{0}^{2} F_{i}(k) }{k R_{i}^{2} } \sin(2kR_{i} + \phi_{i}(k)) e^{(-2\sigma_{i}^{2}k^{2})} e^{\frac{-2R_{i}}{\lambda(k)}},
\end{equation}
where the EXAFS function $\chi(k)$ represents the normalized x-ray absorption oscillations after the atomic absorption background has been removed. Here, $N_i$ is the number of neighbors in a given scattering shell $i$, $R_i$ is the scattering half-path length, and $\sigma_{i}^{2}$ is the atom-pair distance distribution variance, also known as the EXAFS Debye-Waller factor. The other parameters, such as the effective scattering amplitude, $F_i(k)$, effective scattering phase shift, $\phi_i(k)$, and mean free path, $\lambda(k)$, are typically taken from theoretical calculations. Lastly, $S_0^2$ is an overall amplitude reduction factor, typically due to inelastic photoelectron losses. It should be readily apparent that a Fourier transform (FT) of Eq.~\ref{eq:exafs} will produce peaks in $r$-space, and that is how the data will be presented below. Radiation damage can manifest itself in Eq.~\ref{eq:exafs} as changes in any of the structural parameters $N_i$, $R_i$, and/or $\sigma^2_i$.

The amplitude of a peak in the FT of Eq.~\ref{eq:exafs} is highly dependent on $N_i$ and $\sigma_{i}^{2}$. Strictly speaking, when looking at the magnitude of the FT, there can also be interference effects from the real and imaginary components which can affect the magnitude, which is why EXAFS does not produce true radial distribution functions (RDFs). However, in a well ordered system with well separated scattering shells, these interference effects are expected to be minimal for the first few neighbors where multiple scattering does not contribute significantly. As a result of self-irradiation damage there is a decrease of EXAFS amplitude, which corresponds to either a decrease of $N_i$, or an increase in $\sigma_{i}^{2}$. The loss of EXAFS amplitude in the sample is not a result of a homogeneous amorphization of the sample as a whole, but rather from an inhomogeneous combination of undamaged regions and disrupted regions with local environments from within a damage cascade. Although atoms in both the damaged and undamaged regions contribute to the x-ray absorption, the atoms in sufficiently ordered (undamaged) regions dominate the contribution  to the EXAFS oscillations. 

When presenting the results in Sec.~\ref{results}, three different methods are used in characterizing the radiation damage. All three methods consider the $T_A=300$~K annealed data to be the undamaged bulk standard data. This assumption is supported by other experimental annealing studies, including resistivity and magnetic susceptibility measurements that show less than a 2\% change above $T_A=300$~K,\cite{fluss2004temperature} and susceptibility studies that show essentially no change above $T_A = 300$~K.\cite{mccall2006emergent} We have also verified that no change is observed in the EXAFS data for samples annealed at $T_A=375$~K compared to those annealed at room temperature, consistent with thermal expansion measurements.\cite{freibert2007formation} In addition, the sample was annealed at 375~K for about 30 minutes prior to starting the storage phase in the experiments reported here. 

The first method relies only on the raw data, taking ratios of amplitudes of data at a given $T_A$ to those data at $T_A=300$~K. This method should give the damaged fraction of the material in the broadest sense, defining any displacement as damage. The second method relies on conventional EXAFS fitting methods and allows for a separate accounting of moderate disorder. The third method is new to this study, and uses a weighted version of the $T_A=300$~K data to extract the average ordered crystallite radius and the fraction of the material that is so ordered.

\subsection{Amplitude-Ratio (AR) Method}

The most straightforward method simply takes the ratio of the peak amplitude of a given peak in the FT, $A_i$, to that same peak in the FT for the $T_A=300$~K data. The damage fraction, $F_d^\text{AR}$, as a function of $T_A$, is then defined as follows:
\begin{equation}
\label{eq:Fd}F_d^\text{AR} = 1 - \frac{A_i(T_A)}{A_i(T_{300})}.
\end{equation}

We apply this formula in Sec.~\ref{results} to three of the prominent peaks in the FT. The underlying assumption is that radiation damage is strong enough that local distortions from the fcc structure are random enough and large enough ($\sigma^2 \gtrsim 0.04$~\AA$^2$) that such local environments no longer contribute significantly to the EXAFS oscillations. To the extent that $\sigma^2$ is more moderately enhanced in the damaged regions, $A_i \sim 1/\sigma$, and therefore such
regions are still partially included using this method. This method is the primary one used in previous work on the sample studied in this article.\cite{booth2013self}

\subsection{EXAFS Fitting (EF) Method}

This method uses standard EXAFS fitting procedures to determine structural parameters, which were performed using Larch\cite{newville2013larch} with theoretical EXAFS paths created with FEFF9.6.4.\cite{rehr2010parameter,rehr2009ab,rehr2000theoretical}

The fcc structure used to create the FEFF model for the Pu and Ga edges started with a lattice parameter of 4.604~\AA~ for $\approx 4$ at.~\% Ga $\delta$-Pu at less than 50~K.\cite{lawson2014gruneisen,ellinger1964plutonium} The values for the effective scattering amplitude, $F_i(k)$, effective scattering phase shift, $\phi_i(k)$, and mean free path, $\lambda(k)$, were taken from the FEFF calculations.

Since the differences between three scattering shells are already considered in the AR Method results, here we focus on extending those results to the nearest-neighbor Pu-Pu pair distance and Debye-Waller factors. The damage fraction is more clearly defined as only due to atoms in strongly distorted/disordered environments, as the more moderate disordered environments will manifest as enhancements of $\sigma^2$.  To accomplish these comparisons, only the nearest-neighbor shell is reported below (Pu-Pu or Ga-Pu), but a constrained Ga near neighbor peak and an impurity peak near 3.7 \AA~ are also included in the Pu edge fits to account for any peak overlap effects, using the same constraints in the previous study.\cite{booth2013self} More specifically, for the first near neighbor in the Ga K-edge data, a single scattering Ga-Pu path was used, based on the model for fcc $\delta$-Pu, with the lattice parameter allowed to contract to account for the shorter Ga-Pu distance.\cite{booth2013self} This contracted distance was then used in the first shell fit of the Pu $\mathrm{L_{III}}$-edge data, which included a Pu-Ga scattering path with 4.3\% of the number of neighbors in the main Pu-Pu scattering peak at approximately 3.20 \AA. The next near neighbor shell was a Pu-Pu scattering peak at approximately 3.69 \AA~ from a possible impurity phase,\cite{booth2013self} and will be discussed below. Additional paths corresponding to $\alpha^\prime$ phase or PuO$_2$ were added to the fitting to test for possible impurities, however, neither produced physically meaningful fits. The Pu $\mathrm{L_{III}}$-edge data were transformed between 2.5 and 13.24 \AA$^{-1}$ with a Hanning window function with $dk=1$ \AA$^{-1}$, and are fit between 2.5 and 4.1 \AA. The fit utilized 8 parameters and 25 independent data points,\cite{stern1993number} corresponding to 17 degrees of freedom. The Ga $\mathrm{K}$-edge data were transformed between 1.911 and 12.123 \AA$^{-1}$ with a Hanning window function with $dk=1$ \AA$^{-1}$, and are fit between 2.5 and 4.0 \AA. The fitting results of the Ga edge did not depend strongly on the start of the k range, and the lower lower limit was chosen to maximize available independent points. The fit utilized 4 parameters and 12 independent data points,\cite{stern1993number} corresponding to 8 degrees of freedom. Here, the damage fraction is quantified in terms of the ratio of numbers of atoms in the first nearest neighbor scattering shell at a given $T_A$ to those at $T_A=300$ K:
\begin{equation}
\label{eq:EFMFd}F_d^\text{EF} = 1 - \frac{N_1(T_A)}{N_1(T_{300})}.
\end{equation}

\subsection{Spherical-Crystallite (SC) Method}

Enhancements in $\sigma^2$ have been observed in radiation damaged samples of PuCoGa$_5$, where it was conjectured that some moderate disorder should occur near the interface of damaged and undamaged regions.\cite{booth2007self} This situation is similar to the case of a crystalline nanoparticle, where an RDF will only extend to the diameter of the nanoparticle. Here, a model of such behavior in radiation damaged material is presented assuming some shape of the crystalline regions and that otherwise damaged material does not contribute to the EXAFS oscillations. For simplicity of modeling, we consider these crystalline domains to be spherical, although as we discuss in Sec.~\ref{results}, these data are not strongly dependent on the exact shape. In this model, there is an explicit loss of amplitude in addition to the strongly disordered regions outside the nanoparticle crystallites that is due to a loss of coordinating atoms for those near the surface of the crystallite, assuming that the atoms near the center will have a coordination environment much like the bulk material. The EXAFS signal, however, will be an average of all the atoms, and the effect only becomes noticeable as the particle size shrinks to the point where a significant fraction of the atoms are surface-like.

Several models have been proposed to determine the average coordination numbers of nanoparticles in specific configurations, but they rely on knowing the morphology of the particle, and often they utilize only the first near neighbor.\cite{frenkel2001view,benfield1992mean,fritsche1993exact,calvin2003determination} In order to estimate the mean size of the undefected regions of the sample we employ an approach for nanoparticle size determination based upon the technique used by Borowski et al.\cite{borowski1999study,borowski1997size} and Calvin et al.\cite{calvin2005estimating,calvin2003determination} Consider a spherical crystallite of radius $R_c$, with an atom at a distance $R_a$ from the center. A spherical shell around that atom of radius $r$ will partially intersect the surface of the particle when $R_c-R_a \leq r \leq R_c+R_a$. The fraction of that surface enclosed within the particle\cite{calvin2003determination} is \begin{equation}
\label{eq:cal1} \frac{R_c^2-(R_a-r)^2}{4R_a r}.
\end{equation} Integrating over the positions in the particle gives

\begin{multline}
\label{eq:cal_cal1b} \int\limits_{0}^{2\pi}d\phi\int\limits_{0}^{\pi}d\theta\int\limits_{R_c-r}^{R_c} \frac{R_c^2-(R_a - r)^2}{4 R_a r} R_a^2 \sin \theta dR_a d\theta d\phi \\ = \frac{17}{12}\pi r^3 - 4 \pi r^2 R_c + 3\pi r R_c^2 .
\end{multline}

Adding the volume not extending beyond the particle $\frac{4}{3}\pi(R_c-r)^3$, and dividing by the total particle volume gives the change in average coordination number at a given distance and particle size:\cite{calvin2003determination,howell2006pair}

\begin{equation}
\label{eq:cal2}N_{\mathrm{nano}} = \left[ 1-\frac{3}{4}\left(\frac{r}{R_c}\right) + \frac{1}{16}\left(\frac{r}{R_c}\right)^3 \right]N_{\mathrm{bulk}}.
\end{equation}

However, the EXAFS signal in our sample does not come from atoms located in well ordered regions alone. Atoms which have been displaced so significantly as to no longer produce EXAFS oscillations will still contribute to the edge jump of the absorption spectrum, and so during normalization will have the effect of reducing the amplitude of EXAFS and the corresponding FT. To account for this effect, an extra correction factor, developed from Eq.~\ref{eq:cal2}, is utilized,
\begin{equation}
\label{eq:cal3}|\chi_{\mathrm{nano}(r)}| \approx \left[ 1-\frac{3}{4}\left(\frac{r}{R_c}\right) + \frac{1}{16}\left(\frac{r}{R_c}\right)^3 \right] \left( |\chi_{\mathrm{bulk}(r)}| F_c \right),
\end{equation}
where we have generalized and approximated the $N$ parameters to the magnitude of the FT of $\chi(k)$ and $F_c$ represents the fraction of material in the sample left in a crystalline configuration, as opposed to the more distorted/damaged regions outside the crystallites. Non-linear least squares fitting of the RDFs and EXAFS data to Eqns.~\ref{eq:cal2} and \ref{eq:cal3} was performed using Larch.\cite{newville2013larch}

\section{Results}
\label{results}

\begin{figure}
\includegraphics[width=1.0\columnwidth, keepaspectratio=true]{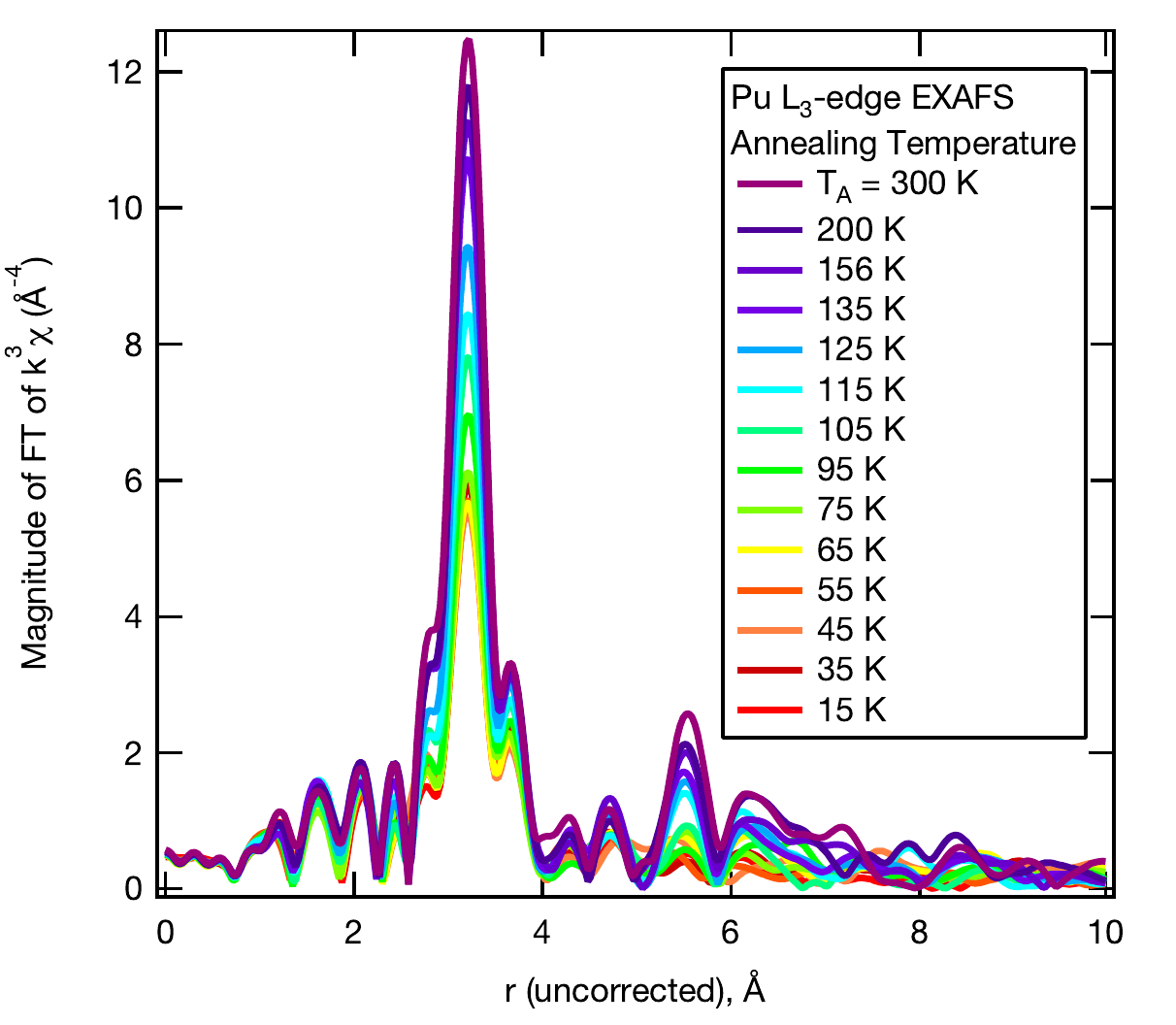}
\caption{\label{fig:purspace} EXAFS Fourier transform (FT) magnitudes at the Pu $\mathrm{L_{III}}$-edge, measured at 15 K, showing the evolution of structure at various $T_A$.}
\end{figure}

\begin{figure}
\includegraphics[width=1.0\columnwidth, keepaspectratio=true]{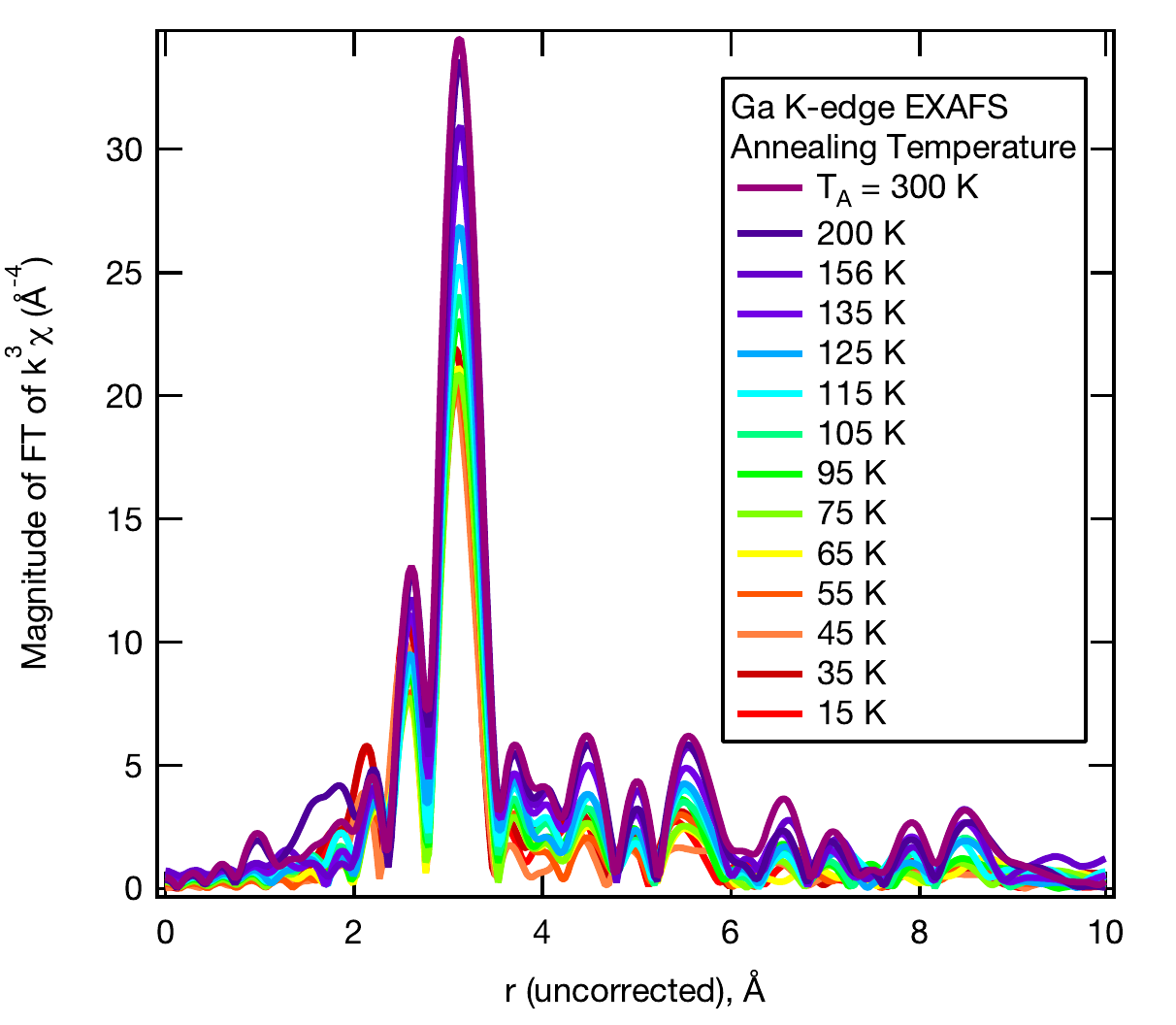}
\caption{\label{fig:garspace} EXAFS FT magnitudes at the Ga K-edge, measured at 15 K, showing the evolution of structure at various $T_A$.}
\end{figure}

The Fourier-transformed EXAFS of the Pu $\mathrm{L_{III}}$-edge and Ga K-edge after each annealing stage are shown in Figs.~\ref{fig:purspace} and \ref{fig:garspace}, respectively. As in the previous study,\cite{booth2013self} the reduction in peak height is attributed to a decrease in $N_i$, as a result of atoms being so displaced from their positions in the lattice that they no longer appear as neighbors.  

To complete and compare to the previous study, these new results are first analyzed using the AR Method, which tracks the reduction of the amplitude of the EXAFS peak compared to its fully annealed value as a function of $T_A$. We then present results of the EF Method to consider more subtle changes in the EXAFS signal with damage and annealing. Finally, the results of the SC Method to determine the crystalline fraction and crystallite radius parameters, $F_c$ and $R_c$, are reported as functions of $T_A$.  

\subsection{Amplitude-Ratio (AR) Method}

\begin{figure}
\includegraphics[width=1.0\columnwidth, keepaspectratio=true]{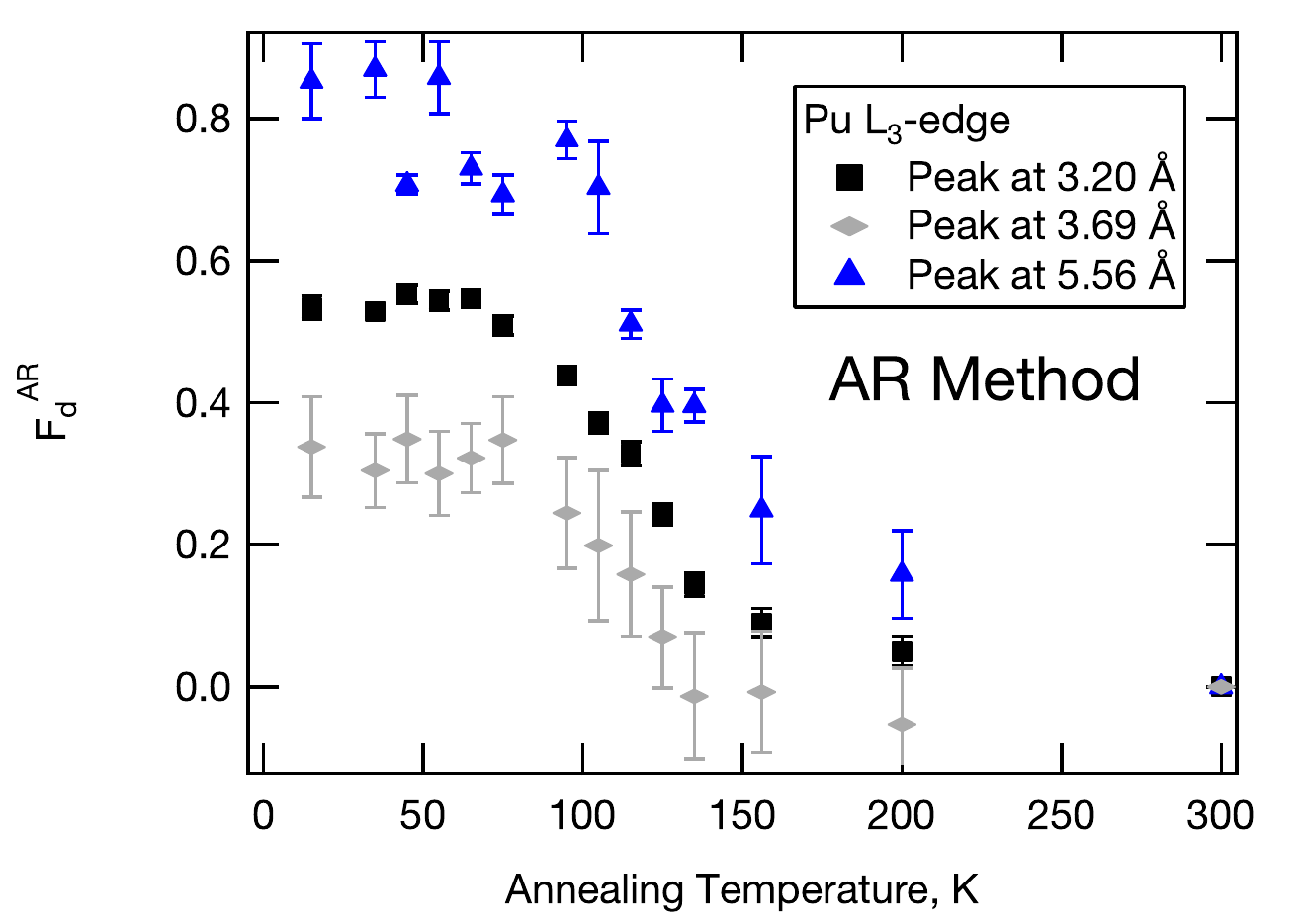}
\caption{\label{fig:puheightFd} Damage fraction of selected peaks in the Pu $\mathrm{L_{III}}$-edge EXAFS as determined by the amplitude ratio method relative to fully annealed intensity, as function of $T_A$.}
\end{figure}

\begin{figure}
\includegraphics[width=1.0\columnwidth, keepaspectratio=true]{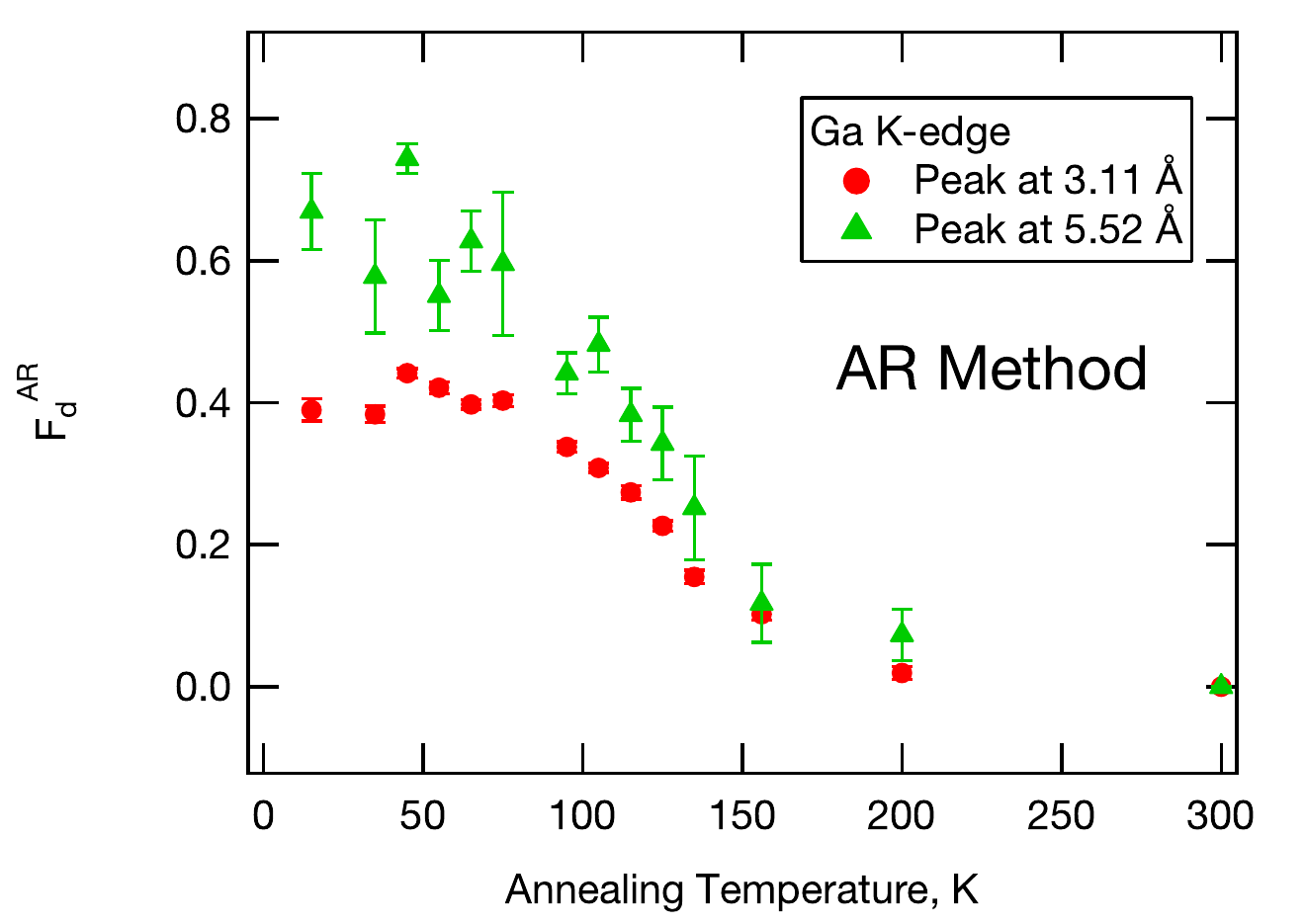}
\caption{\label{fig:gaheightFd} Damage of selected peaks in the Ga K-edge EXAFS as determined by the amplitude ratio method relative to fully annealed intensity, as function of $T_A$.}
\end{figure}

As mentioned above, this method explicitly ignores more moderate disorder (eg. $\sigma^2$ enhancements less than $\approx$0.04~\AA$^2$), but implicitly includes such disorder fractionally to the extent it affects the EXAFS amplitudes ($A\sim1/\sigma$). The damaged fraction of selected peaks in the Pu $\mathrm{L_{III}}$-edge and Ga K-edge EXAFS as a function of $T_A$ are shown in Figs.~\ref{fig:puheightFd} and \ref{fig:gaheightFd}, respectively.

For Pu, Fig.~\ref{fig:puheightFd} shows close to 60\% of the Pu atoms have become heavily disordered, with respect to their first nearest neighbors. For the second nearest neighbors this number jumps to above 80\%. The large error bars for the second shell amplitudes at the lowest temperatures are a result of the height of that peak being diminished close to the point of experimental noise. As in the previous measurements on this sample,\cite{booth2013self} we see a non-fcc impurity component to the EXAFS signal evident as a Pu-Pu pair distance of approximately 3.69 \AA, between the first and second shells in $\delta$-Pu. This feature indicates less damage accumulation than indicated in the first shell, despite being farther away, consistent with it being due to a distinct phase from $\delta$-Pu. Since the contribution of the impurity peak to the overall signal is small, further structural identification is difficult, but its low concentration in the sample should not significantly affect the other results presented here.  This peak will be discussed further in Sec.~\ref{disc}.

The Ga K-edge data, Fig.~\ref{fig:gaheightFd}, show a similar trend; however, we note the overall amount of damage is not as high, with only 40\% in the first shell, and close to 60\% in the second shell. The tendency for Ga to hold on to more of its neighbors may be evidence for its efficacy as a $\delta$ phase stabilizer consistent with Ref.~\onlinecite{allen2002vibrational} and with previous measurements showing little or no radiation damage around Ga compared to Pu in room temperature annealed material.\cite{booth2013self} It is important to note, however, that this decreased damage around Ga is in contrast to the previous experiment on the same sample but with the damage recorded at $T_A=30$~K, where the Ga edge amplitude was similar to, but slightly higher than, that recorded from the Pu edge.\cite{booth2013self} This variability is consistent with the changes noted by Conradson et al.\cite{conradson2014nanoscale,conradson2014intrinsic} leading to the conclusion that exact sample history can affect the EXAFS signal.

\begin{figure}
\includegraphics[width=1.0\columnwidth, keepaspectratio=true]{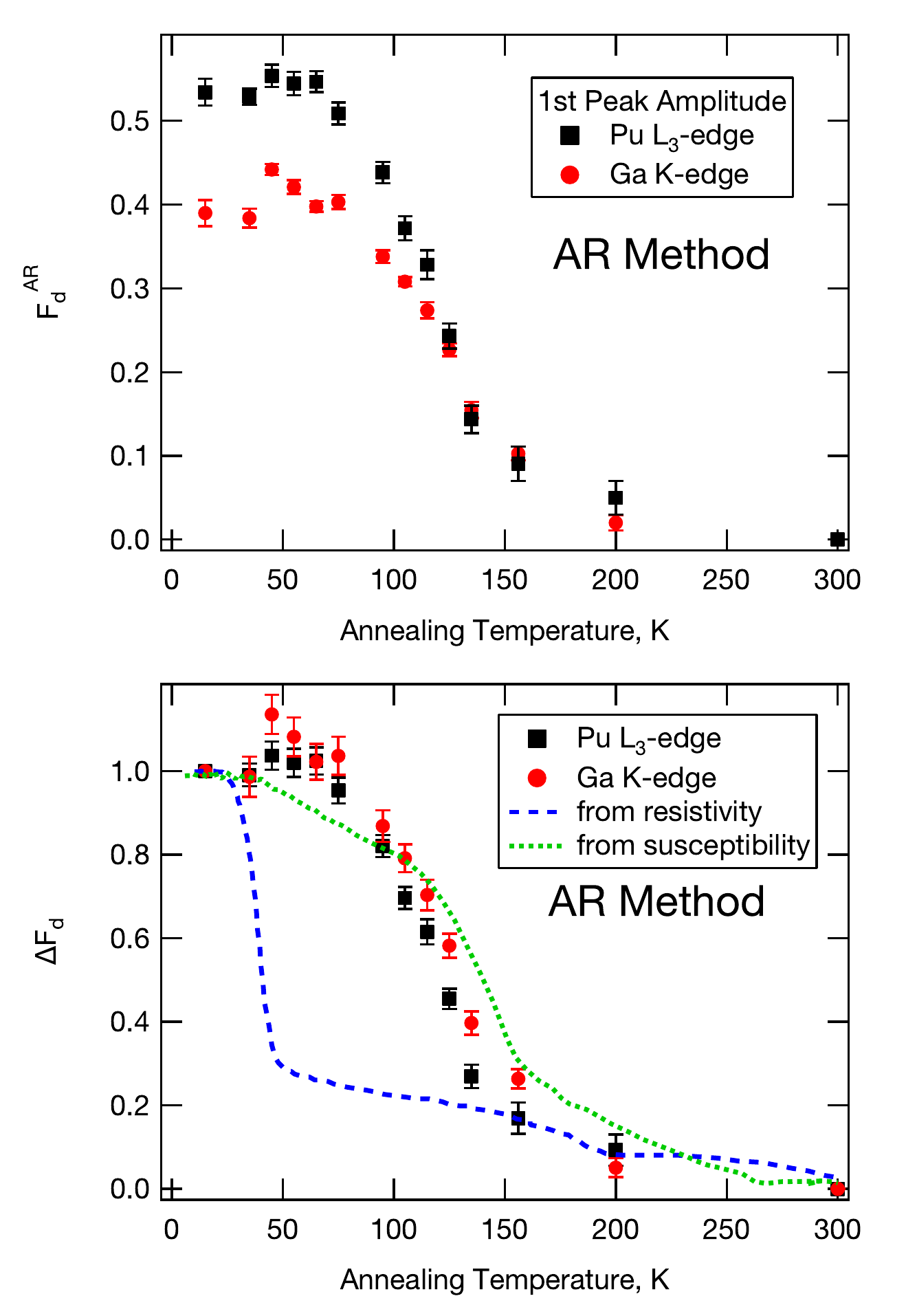}
\caption{\label{fig:dFdPeakHeight} Damage to local structure of $\delta$-Pu after cold storage for 72 days below 15 K as determined by peak amplitude (top). The data are replotted to show the normalized fraction of residual damage present as a function of $T_A$ (bottom) along with similarly normalized data from previous resistivity\cite{fluss2004temperature} and susceptibility\cite{mccall2006emergent} isochronal annealing experiments.}
\end{figure}

For comparison to prior EXAFS experiments\cite{booth2013self} as well as previous resistivity\cite{fluss2004temperature} and susceptibility\cite{mccall2006emergent} measurements, in several of the following graphs, we report the fractional change in damage $\Delta F_d$ as measured by a given quantity $Q$ (eg. resistivity or EXAFS amplitude change) as a function of $T_A$, 
\begin{equation}
\label{eq:dFd}\Delta F_d = \frac{Q_{T_{300 \mathrm{K}}}-Q_{T_A}}{Q_{T_{300\mathrm{K}}}-Q_{T_{15\mathrm{K}}}},
\end{equation}
where $Q_{T_{15}}$ is the value of that quantity prior to any annealing. The fractional change in damage determined by the amplitude ratio method for first near neighbors of Pu and Ga are shown in Fig.~\ref{fig:dFdPeakHeight}.

\subsection{EXAFS Fitting (EF) Method}

\begin{figure}
\includegraphics[width=1.0\columnwidth, keepaspectratio=true]{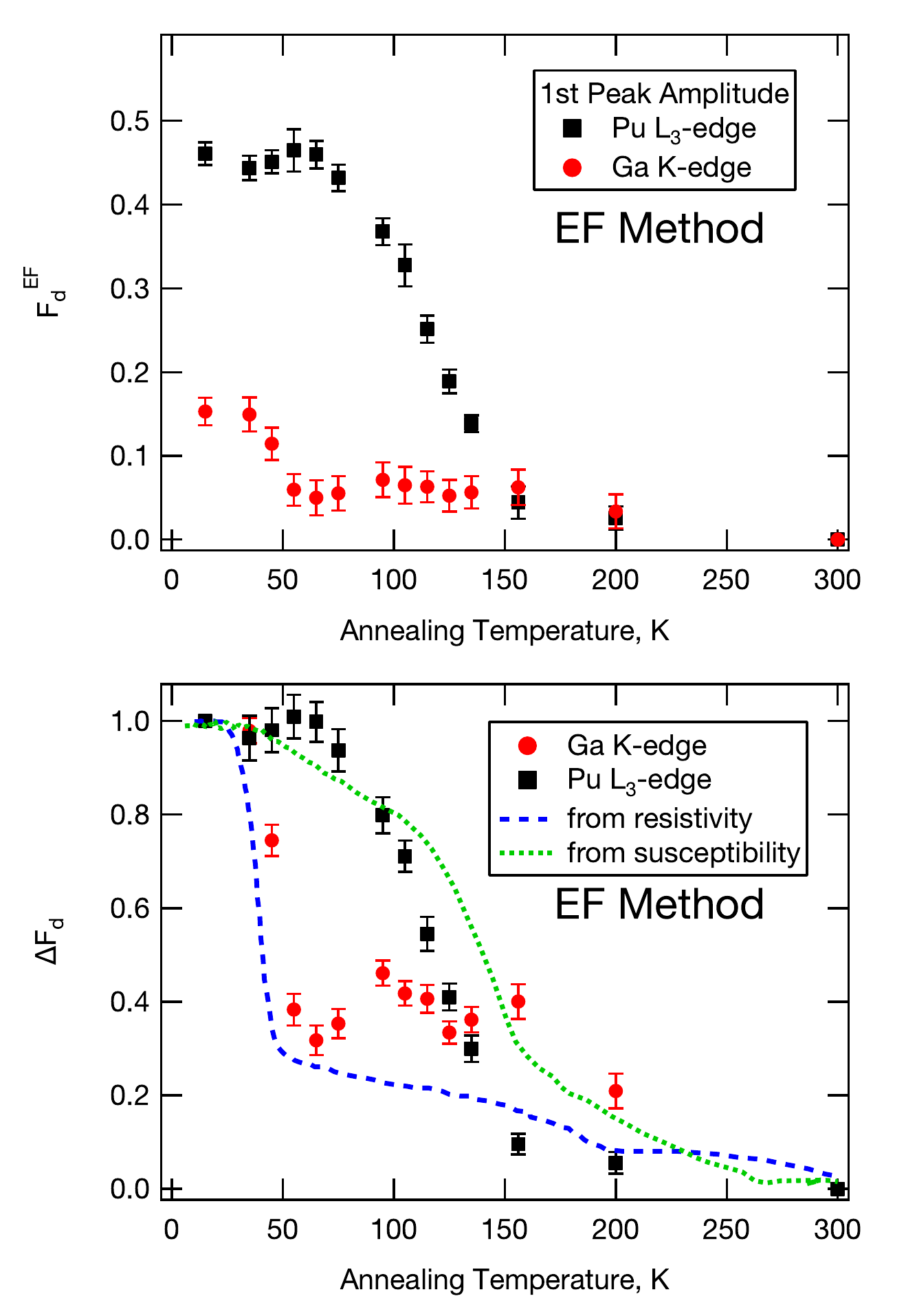}
\caption{\label{fig:dFdFitting} Damage to local structure of $\delta$-Pu after cold storage for 72 days below 15 K as determined by EXAFS fitting of peak amplitude (top). The data are replotted to show the normalized fraction of residual damage present as a function of $T_A$ (bottom) along with similarly normalized data from previous resistivity\cite{fluss2004temperature} and susceptibility\cite{mccall2006emergent} isochronal annealing experiments.
}
\end{figure}

\begin{figure}
\includegraphics[width=1.0\columnwidth, keepaspectratio=true]{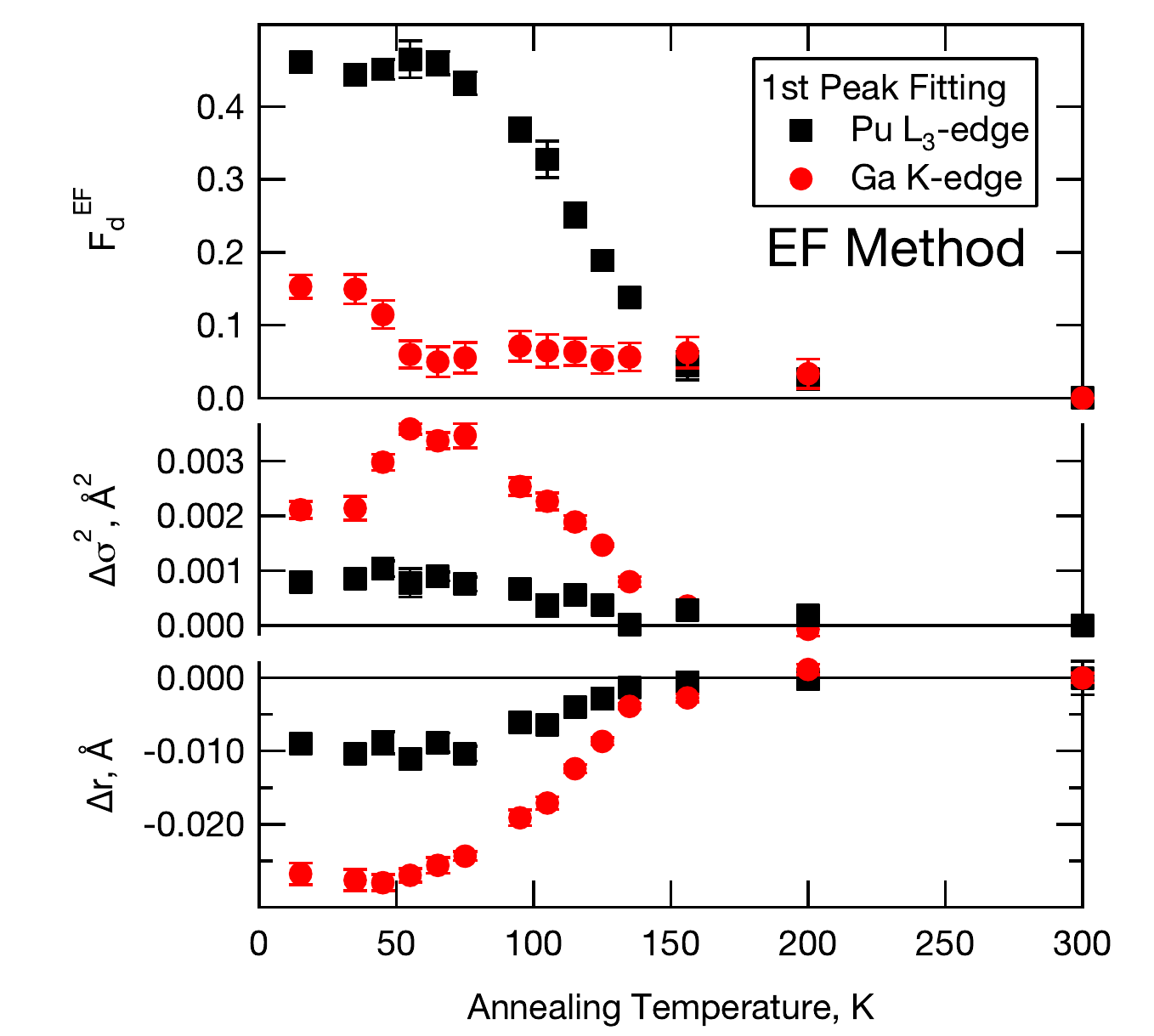}
\caption{\label{fig:FdLTssdisdelr} Damage fraction as determined by EXAFS fitting of peak amplitude (top), along with additional disorder (middle) and bond length (bottom) relative to fully annealed data required for best fit.}
\end{figure}

By examining the data in more detail with a full EXAFS structural model, we can account for both total loss of EXAFS signal from Pu and Ga atoms in highly disordered/distorted environments through overall losses of amplitude, and more moderate disorder from enhancements to $\sigma^2$ from atom pairs experiencing static (non-thermal) disorder less than about 0.04 \AA$^2$, such as may occur near the surface of an ordered particle crystallite or in the neighborhood of a defect. $F_d^\text{EF}$ and $\Delta F_d^\text{EF}$ using $N_i$ and Eqs. \ref{eq:EFMFd} and \ref{eq:dFd} from EXAFS fits are shown in Fig.~\ref{fig:dFdFitting}, together with previous resistivity\cite{fluss2004temperature} and susceptibility\cite{mccall2006emergent} measurements for comparison. The additional disorder and change in scattering distance relative to the fully annealed data are plotted in Fig.~\ref{fig:FdLTssdisdelr}. 

By separately accounting for the moderately disordered atoms, we see that the damaged fraction as measured by the overall reduction in the number of neighbors compared to the amplitude reduction of the first peak in the FT of the Pu EXAFS drops slightly to just above 40\%, but the fraction of damaged Ga drops by more than half to below 20\%. For Pu the overall shape of the annealing curve still matches the susceptibility measurements, however, for the Ga curve a sharp drop at lower $T_A$, much like in the resistivity curve is evident, Fig.~\ref{fig:dFdFitting}.

Above the $T_A=135$ K, where the amplitudes have mostly recovered and the damage fraction is below around 10\%, very little additional disorder or change in bond length is observed for either Pu-Pu or Ga-Pu. Below that temperature, however, their behaviors are quite different. In terms of extra disorder, the Pu-Pu scattering path requires approximately an extra 0.01 \AA$^2$ from $T_A=15$~K to $T_A=95$~K, which then decreases, approaching zero static disorder above 150~K. The Ga-Pu path requires a similar amount of additional disorder at the lowest annealing temperature, but increases to more than double the amount at 55 K before starting to decline again at $T_A=95$ K. We must emphasize this observation because it indicates that the Ga-Pu environment becomes more disordered between $T_A = 55$~K and 150~K than for $T_A$ below 50~K; that is, the Ga environment is more ordered at temperatures where more of the sample is damaged (according to the Pu edge results from the far more numerous Pu atoms) than at temperatures where that damage is being removed by annealing.

Below $T_A=135$ K, the EXAFS data from both the Ga and Pu edges fit best with a short bond length contraction, although the effect is smaller for Pu-Pu, less than 0.01 \AA, compared to the almost 0.03 \AA~ contraction seen from the Ga-Pu path.

\subsection{Spherical-Crystallite (SC) Method}

\begin{figure}
\includegraphics[width=1.0\columnwidth, keepaspectratio=true]{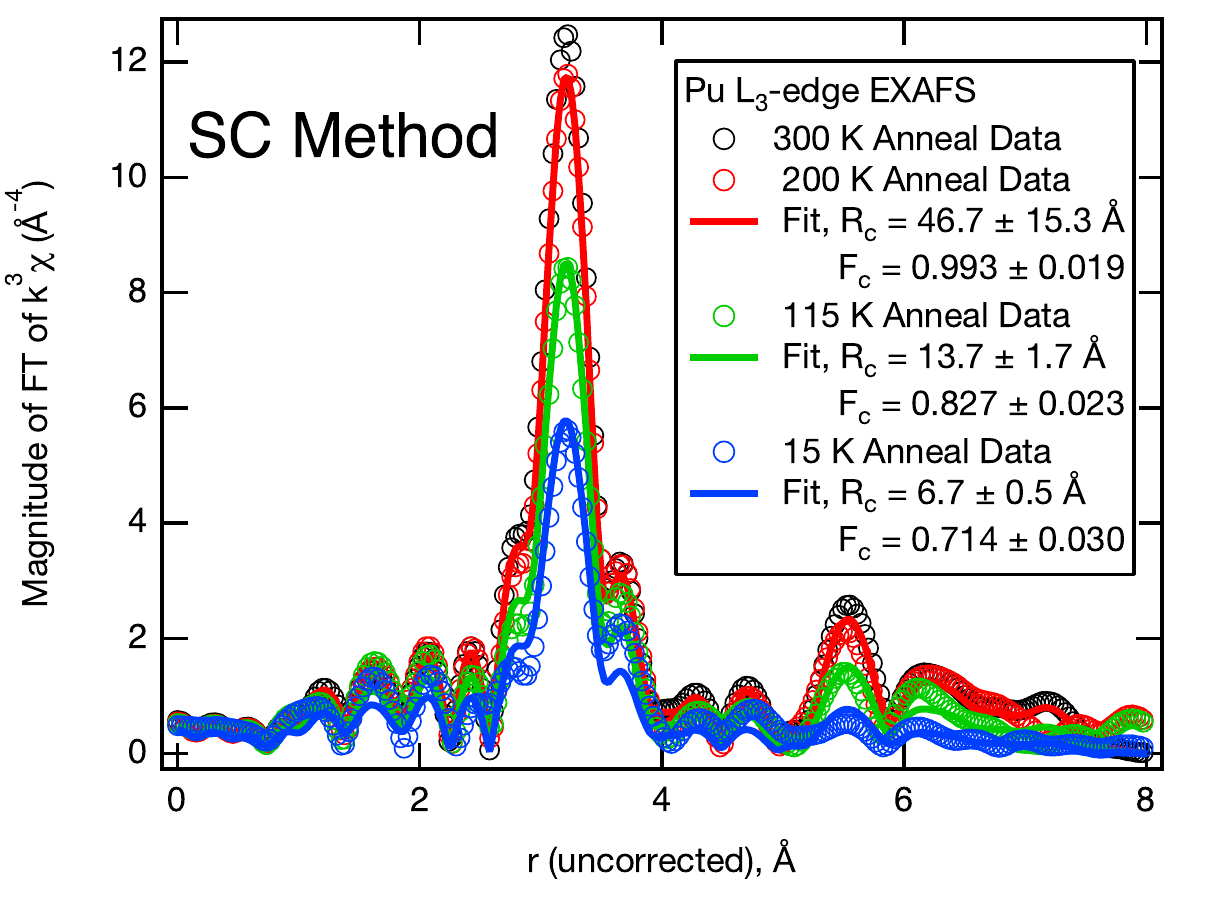}
\caption{\label{fig:pu_calfit} Fitting of Pu $\mathrm{L_{III}}$-edge data at various stages in the annealing process to the fully annealed $T_A=300$ data, using Eq.~\ref{eq:cal3}.}
\end{figure}

\begin{figure}
\includegraphics[width=1.0\columnwidth, keepaspectratio=true]{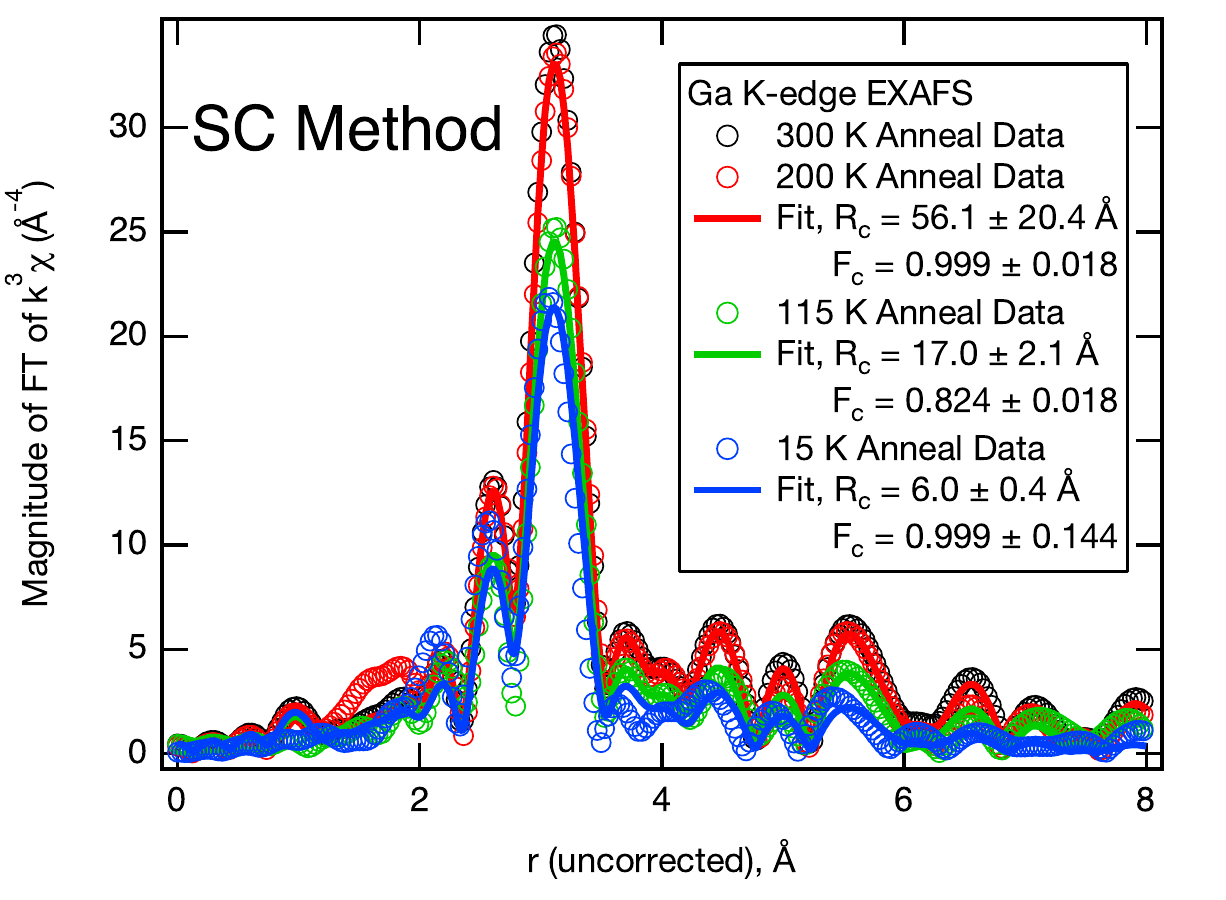}
\caption{\label{fig:ga_calfit} Fitting of Ga K-edge data at various stages in the annealing process to the fully annealed $T_A=300$ data, using Eq.~\ref{eq:cal3}.}
\end{figure}

\begin{figure}
\includegraphics[width=1.0\columnwidth, keepaspectratio=true]{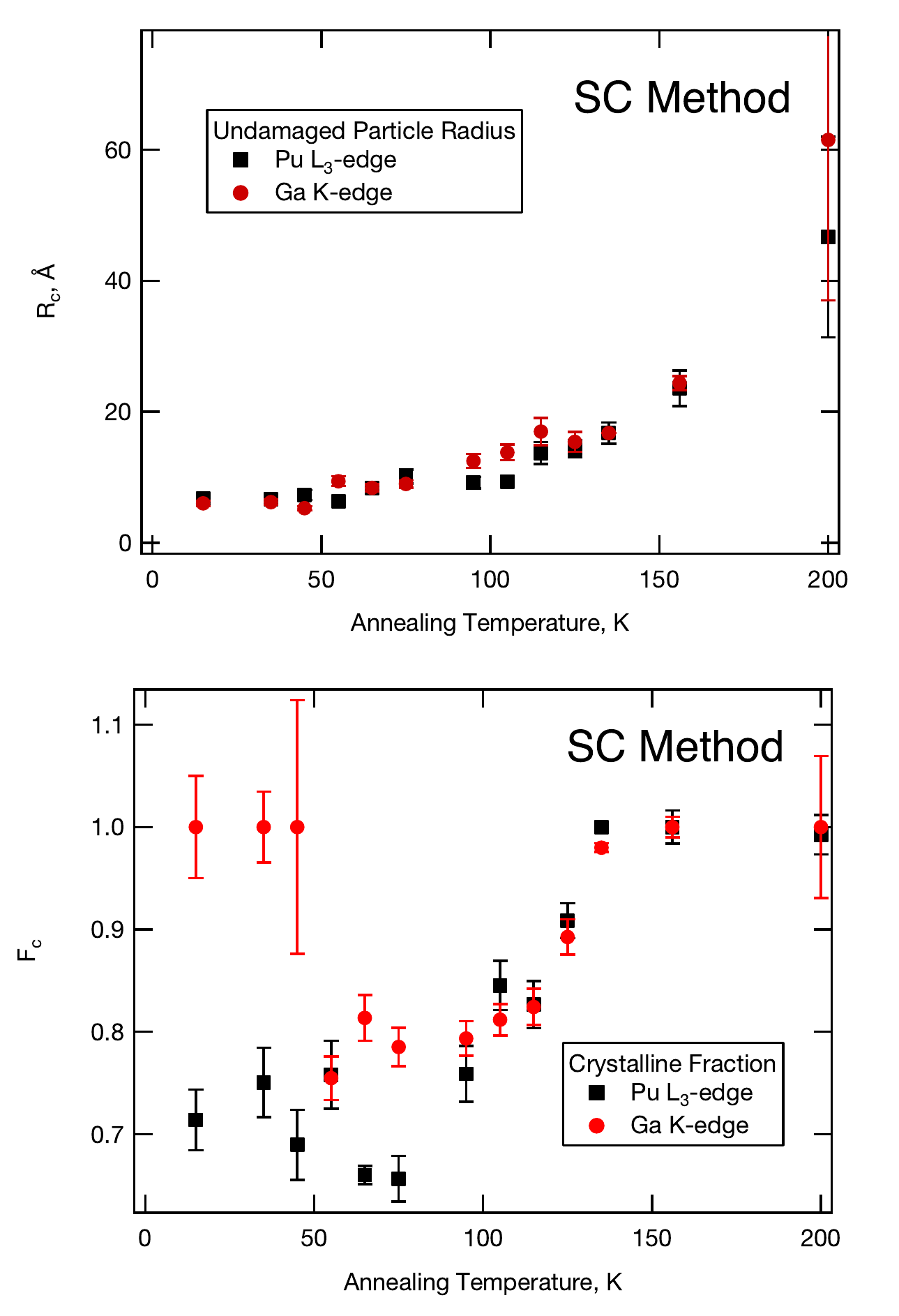}
\caption{\label{fig:calfits} (upper) Average radius $R_c$ of the finite-sized crystallite regions and (lower) the crystalline fraction $F_c$ as a function $T_A$, as determined by fitting the Pu $\mathrm{L_{III}}$-edge and Ga K-edge data using Eq.~\ref{eq:cal3}.}
\end{figure}

Treating the data with the SC Method of Eq.~\ref{eq:cal3} allows for the use of information contained in the farther scattering paths without performing traditional EXAFS fits and takes account of surface-to-volume effects at the edges of ordered regions of the material. Examples of fitting to data at selected $T_A$ for the Pu and Ga edges are shown in Figs.~\ref{fig:pu_calfit} and \ref{fig:ga_calfit} respectively. Results for $R_c$ and $F_c$ at all measured $T_A$ are plotted in Fig.~\ref{fig:calfits}.

Although there is strong agreement between the data and the SC model,  difficulties fitting the data with this simple model arise close to the extremes of particle radius. For example, at the very largest sizes, there is little difference between an extremely large particle in a partly amorphous matrix, or smaller particles in a mostly crystalline matrix. Likewise at the other end of the scale, with a particle so small as to have almost no long distance peaks, it becomes difficult to differentiate between loss of amplitude as a result of termination effects and loss due to amorphous material. The only other problematic data are the Pu edge data at 75 K and the Ga edge data at 135 K, which produced inconsistent values of $F_c$ and $R_c$ with large error bars. Here, we have chosen to constrain the Pu particle radius to the Ga value at 75 K and the Ga radius to the Pu value at 135 K, since the $R_c$ are generally very consistent over the entire temperature range.

Of particular note, values of $R_c$ measured from both edges are remarkably consistent, with the Ga radius higher than Pu at the lowest temperatures,  $6.0 \pm 0.1$~\AA~ versus $4.3 \pm 0.06$~\AA. This difference may be an artifact of the lower crystallinity in Pu at these temperatures.  The same is generally true of the $F_c$ measurements, although one other feature stands out: For $T_A < 50$~K, $F_c$ from the Ga edge is clearly larger than from the Pu edge, a situation which is directly reflected and supported by the $\sigma^2$ behavior in Fig.~\ref{fig:FdLTssdisdelr}. The anticorrelation between $F_d^{EF}$ and $\sigma^2$ in the Ga edge data is better clarified using the SC model, in which the increase in $\sigma^2$ can be attributed to a drop in crystallinity as the damage in the Ga atoms is annealed.

Considering the use of this method for the study of radiation damage is somewhat novel, it is worth discussing the limitations of the model. When used with discrete nanoparticles for size determination, this method typically produces results that are smaller than found by Scherrer analysis of x-ray diffraction (XRD) data,\cite{calvin2003determination,calvin2005comparison,calais1998crystallite} an effect attributed to XRD being more sensitive to the size distribution of larger particles compared to EXAFS which is more sensitive to smaller particle sizes in a polydisperse sample.\cite{calvin2005estimating} Transmission electron microscopy (TEM) has also been shown to produce estimates of particle size larger than EXAFS for discrete particles.\cite{calvin2005comparison,frenkel1999solving} In addition, TEM estimates larger defect free regions, based on dislocation loop size and density, in radiation damaged bulk material.\cite{gupta2015radiation,olive2016using} The inability of TEM to capture the smallest sized defects in a material, and the sensitivity of EXAFS to the increase in disorder due to relaxation effects around various defects have been proposed as a reason for the discrepancy.\cite{calvin2005comparison,frenkel1999solving}

While this method assumes a spherical shape to the ordered nanoparticle crystallites, the model fits the data well enough that attempts to fit to different shapes did not produce meaningful results. In particular, an ellipsoidal lineshape was used, but correlations between the added parameters  made finding stable fits which were physically meaningful difficult. Indeed, it is remarkable that the SC model is even applicable to this system. For example, Li et al.\cite{li2012study} found that Mo, which is substitutional in Fe like Ga is in $\delta$-Pu, only showed damage in its first two coordination shells between 0 and 1 dpa of neutron irradiation, and further damage only occurred in the first coordination shell out to 10 dpa. Thus the spherical damage model would not have applied in that case. Although at 10 dpa those results are well beyond the percolation limit, this shows that despite its simplicity, it is far from a universal model that can be made to fit any kind of radiation damage, and the fact that it can be used to fit both Ga and Pu edge data in this case is worth noting.

\section{Discussion}
\label{disc}

There are several important features of the results presented above. First, the vastly enhanced data quality due to the ability to measure all the data at 15~K over the previous work where the data were collected at $T=T_A$ has indeed allowed for a clearer interpretation of the changes as due to structural reorganization as opposed to vibrational effects. Together with the increased number of temperature data points, particularly in the critical 50~K to 140~K range, allow for a more precise discussion when comparing to other data and the various annealing stages.

Before discussing the results as a whole, we point out that one of the more striking results presented above is the agreement between the data and the SC model, despite the fact that it is not immediately obvious how recoil nuclei leaving trails of damage throughout the material can leave the undamaged regions, on average, in a configuration that is consistent with a spherical crystallite. It is, however, clear that the strong amount of damage is generally consistent with the magnetic susceptibility\cite{mccall2006emergent} and
previous EXAFS results\cite{booth2007self,booth2013self} that suggest that, in addition to the damage as measured by defect concentrations, all the atoms within a damage cascade experience a significant distortion. This result is also consistent with that expected from the displacement fields generated by the various defect structures. For instance, the presence of an interstitial atom in the lattice will distort its neighbors away from their equilibrium positions. For a tetrahedral self-interstitial in fcc iron, a displacement field $\geq$ 1.5\% of the lattice constant has been calculated to extend throughout 64 atomic volumes.\cite{beeler2012radiation} It therefore seems plausible, whether due to recrystallization into new microcrystals at the end of a displacement spike, or because of the formation of a region surrounded by interstitials and their displacement fields, the crystal structure might be broken into regions that are consistent with a spherical particle interpretation.

The various methods of characterizing radiation damage are complementary. Looking at the two methods that concentrate on the nearest-neighbors for determining the fraction of material that has been damaged, $F_d^\text{AR}$ is the most encompassing and should give the largest estimate. Indeed, $F_d^\text{AR}$ is about 53\% and 39\% at temperatures below 55~K from the Pu and Ga edges (Fig.~\ref{fig:dFdPeakHeight}), while $F_d^\text{EF}$ is about 46\% and 15\% (Fig.~\ref{fig:dFdFitting}), respectively. The decrease is due to separately accounting for moderate disorder in the EF Method, and indeed, the large difference from the Ga edge data is mirrored by the generally large differences
in $\sigma^2$ and $R$ for the nearest neighbors.

These differences can be better understood when comparing to the results of the SC Method. In fact, the most striking result reported here is the high $F_c$ for the Ga data compared to the Pu data below 50~K (Fig.~\ref{fig:calfits}). Taken together with the low $\sigma^2$ for the Ga-Pu pairs in the same temperature range, we can understand the large difference for the Ga edge data between $F_d^\text{AR}$ and $F_d^\text{EF}$ as a partial consequence of not properly considering finite particle-size effects in those methods. This conclusion is consistent with the observed bond length contractions (Fig.~\ref{fig:FdLTssdisdelr}) in the crystallite regions as they become smaller (Fig.~\ref{fig:calfits}), an effect observed in other nanoparticle systems.\cite{huang2008coordination,gilbert2004nanoparticles,frenkel2001view} It is interesting to note that the Ga-Pu bond length contracts more than the Pu-Pu bond length with decreasing particle size, and may be indicative of the average reorganized species in the ordered crystallites at the lowest $T_A$, although we point out that, for instance, such a bond length contraction is not expected from the structure of Pu$_3$Ga.\cite{ellinger1964plutonium}

It is instructive to consider these data together with the different annealing stages mentioned in Sec.~\ref{intro}. The temperature ranges for these stages were identified by Fluss et al.\cite{fluss2004temperature} using the isochronal resistivity data. According to that work, Stage I occurs between 0~K and 45~K, Stage II between 45~K and 110~K, Stage III between 110~K and 180~K, Stage IV between 180~K and 310~K and Stage V above 310~K. These temperatures are consistent with those observed in the magnetic susceptibility work.\cite{mccall2007isochronal}

The physical picture that emerges from these results is that within Stage I annealing below 50~K, Ga atoms form local crystalline structures (high $F_c$ from Ga edge). Given that such a large fraction of the material has been damaged (high $F_d$s from the Pu edge in general), these structures must include a significant number of Ga atoms that are within the volume of the damage cascades. Since $T_A$ is low, they must form during the thermal-spike phase and these crystallite structures quench before they can diffuse into the main structure. When the material enters into Stage II annealing above 50~K, the Ga can diffuse into the rest of the structure, and thereby take on structural properties more similar to the average system structure, as exemplified by the Pu edge data, and therefore having $F_c$ values more similar between the two absorption edges above 50~K (Fig.~\ref{fig:calfits}).

This picture fits well with that derived from other studies, both experimental and theoretical. Beginning with experiment, resistivity\cite{fluss2004temperature} and magnetic susceptibility\cite{mccall2006emergent} isochronal annealing experiments were performed on cryogenically aged $\delta$-Pu, and the fractional change in damage measured by those experiments is plotted along with data from this experiment. Resistivity depends linearly on defect concentration, and in fact, such experiments have been used in the past to estimate defect concentrations.\cite{averback1978ion} Magnetic susceptibility, on the other hand, is induced in plutonium as a result of localized magnetic moments created by perturbations to the crystal structure, which heal and return the $\delta$-Pu to its normal nonmagnetic state near room temperature.\cite{mccall2006emergent} The isochronal susceptibility experiment was closer in experimental parameters to the present one as the aged sample has the same Ga concentration  and was aged for 42 days at $T \leq 30$ K.\cite{mccall2006emergent} The resistivity experiment aged a sample with lower Ga concentration, 3.3 at.~\%, for 3 days at $T=20$ K.\cite{fluss2004temperature} Although the resistivity curve plotted here for comparison was from samples held at the lowest temperature for a shorter duration to ensure the damaged areas were in the dilute limit,\cite{fluss2004temperature} $\delta$-Pu held for longer times (27.7, and 38.3 days at 4.5 K) exhibited similar overall damage characteristics.\cite{wigley1965effect} Those samples also used higher concentrations (8 at.~\% and 4 at.~\%, respectively) of a different $\delta$ stabilizer, Al, suggesting that the attributes of the resistivity curve are sufficiently general for comparing to the local structure results presented here. Given that resistivity is fundamentally a transport measurement, care is warranted in interpreting defect population results from a material whose undamaged stage already contains resistive defect sites in the form of Ga atoms.

Interestingly, the susceptibility exhibits its largest recovery at higher temperatures than the resistivity curve, largely in Stage III\cite{mccall2006emergent,schilling1973recovery} where vacancy migrations can occur and annihilate remaining interstitials after near neighbor vacancy/interstitial annihilation occurs in Stage I, while the main change in the resistivity curve occurs at lower temperatures in Stage I near the border with Stage II,\cite{schilling1973recovery} perhaps by rearrangement of interstitials. By filling in the part of the annealing curve missing in the earlier EXAFS work on this sample,\cite{booth2013self} it is evident that the changes measured by EXAFS and the near neighbor AR Method (Fig.~\ref{fig:dFdPeakHeight}) most closely resemble the annealing curve from magnetic susceptibility measurements. This agreement is consistent with the fact that the EXAFS results are directly related to the volume fraction and the conjecture\cite{mccall2006emergent} that the susceptibility is also most dependent on the volume of damaged material. However, results from the Ga edge using the EF Method (Fig.~\ref{fig:dFdFitting}) most closely resemble the annealing curve from resistivity measurements, suggesting that when properly accounting for the effects of moderate disorder, the Ga atoms are responsible for the bulk of residual resistivity in self-irradiation damaged samples. Furthermore, we can infer from these results that the lack of recovery at the Stage I/Stage II border in proton irradiated samples\cite{fluss2004temperature} indicates the configuration of Ga atoms associated with the displacement spike from the U recoil nucleus, and not the $\alpha$ particle, may be responsible for those changes in resistivity in self-irradiated samples. 

The most straightforward comparisons of these results to theoretical calculations are to those of molecular dynamics simulations of collision cascades produced by $\alpha$-decay by Kubota et al.\cite{kubota2007collision} Among many other results, they find that the initial damage structure produced is an amorphous configuration, with a slight contraction in the first near neighbor distance.\cite{noteRD1} After 70~ps at 300~K the defect configuration reached a steady state recovery, into a glass-like pair correlation. This configuration persists for the remainder of the simulation time, up to 2~ns,  in stark contrast to simulations of other fcc metals which reach an annealed configuration in less than 100~ps.\cite{gao1998effects} By turning off the effects of the 5$f$ electrons in their MEAM potential, they were able to produce a defect recovery similar to other fcc metals. At a simulation temperature of 180~K, the cascade spread throughout the entire simulation cell of 2,048,000 atoms (corresponding to a cube side length of about 37~nm). During the time in cold storage, assuming no overlap, there are about 154,000 atoms per decay (the inverse of the age in decays per atom), and only about half of these are removed from the EXAFS signal, suggesting either the damage cascade does not propagate as far as the simulations predict at cold temperatures, or that perhaps there is additional re-ordering of the structure that occurs as it cools that is not reproduced in the simulation. Nevertheless, our results are consistent with the simulation results showing $\delta$-Pu does not behave like other fcc metals. 

Finally, it is interesting to consider the impurity (non-fcc) peak corresponding to Pu-Pu scattering at 3.69 \AA, which is conjectured to be due to a so-called ``$\sigma$-phase.''\cite{conradson2014intrinsic,conradson2014nanoscale} This impurity has been observed by multiple investigators\cite{booth2013self} including pair-distribution function analysis of x-ray diffraction data.\cite{platteau2009pdf} While this peak clearly follows a similar damage dependence on $T_A$ (Fig.~\ref{fig:puheightFd}), it does not appear to be simply related to the finite-sized crystallites, since it shows less absolute damage at the lowest $T_A$s than the nearest-neighbor Pu-Pu pairs, and therefore does not fit the SC model. Solving the role and identity of the $\sigma$-phase here is beyond the scope of this study, but will require an even more careful comparison between samples with the same histories owing to the history dependence observed here (difference relative damage fractions measured from the Ga and Pu edge between this study and the previous one\cite{booth2013self}) and in many other measurements.\cite{conradson2014intrinsic,conradson2014nanoscale} It should be noted that the impurity is not likely to be on the surface of the sample. Considering sample geometry and the absorption length of Pu fluorescence x-rays, the Pu EXAFS signal comes from at least 5 $\mu$m of the sample.

\section{Conclusions}

We have demonstrated several EXAFS analysis methods for interpreting the changes in local structures around Ga and Pu in $\delta$-Pu along various parts of an isochronal annealing curve, taken after 72 days of storage below 15 K.  We demonstrate that a new model for the determination of particle size is applicable to this system. Of particular note, the combination of the methods for characterizing radiation damage effects from the EXAFS measurements demonstrates that Debye-Waller enhancements of the Ga-Pu nearest-neighbor pairs and the estimated crystalline fraction of the material indicate that the Ga environment is actually more ordered at annealing temperatures below 50~K than above, suggesting that Ga attempts to form locally ordered structures during the initial damage event, on the order of $10^{-11}$ s when the material is essentially melted by the deposited energy,\cite{brinkman1954nature} and this order is quenched into the material, and may be responsible for most of the residual resistivity observed in other experiments. Above 50~K, the Ga diffuses\cite{schilling1973recovery} into the main damaged Pu matrix and the Ga local environment takes on a more average structure. The increase of disorder seen in the Ga data at these temperatures, combined with a drop in crystallinity is evidence of the role Ga plays restoring order to the damaged lattice. 

Future isochronal annealing studies should be conducted, perhaps with different soak times to permit a Meechan-Brinkman analysis\cite{meechan1956electrical,wycisk1976quenching} to extract the activation energies of the various annealing processes observable by EXAFS. Direct experimental measurement of these activation energies would make for an excellent comparison to first principles electronic structure based models of damage structures in $\delta$-Pu.

\begin{acknowledgments}
We gratefully acknowledge the support of the U.S. Department of Energy (DOE) through the Los Alamos National Laboratory (LANL) LDRD Program and the Glenn T. Seaborg Institute for Transactinium Science. Work at Lawrence Berkeley National
Laboratory was partially supported by the Director, Office of Science (OS),
Office of Basic Energy Sciences (OBES),
Chemical Sciences, Geosciences, and Biosciences Division of the prepared under U.S. DOE 
Contract No. DE--AC02--05CH11231.
LANL is operated by Los Alamos National Security, LLC, for the National Nuclear Security Administration of the U.S. DOE under Contract DE--AC52--06NA25396. Use of the Stanford Synchrotron Radiation Lightsource, SLAC National Accelerator Laboratory, is supported by the U.S. DOE, Office of Science, Office of Basic Energy Sciences under Contract No. DE--AC02--76SF00515. Work at Lawerence Livermore National Laboratory 
was prepared under DOE Contract DE--AC52--07NA27344.
\end{acknowledgments}

\section*{References}
\bibliography{bib1link}

\begin{thebibliography}{71}
\bibitem [{not()}]{noteRD1}%
  \BibitemOpen
  \href@noop {} {}\bibinfo {note} {Note: This last point isn't quantified in
  their text, although it can be seen in their Figure 11.}\BibitemShut {Stop}%
\end{thebibliography}

\end{document}